\documentclass[aps,prl,showkeys,preprint,groupedaddress]{revtex4}

\usepackage{graphicx, bm, amsfonts, amssymb, amsmath}

\begin{document}

\title{Optimized replica gas estimation of absolute integrals and partition functions}

\author{David D. L. Minh}
\email[Electronic Address: ]{daveminh@gmail.com}

\affiliation{Biosciences Division, Argonne National Laboratory, 9700 S. Cass Av., Argonne, IL 60439 USA}

\date{\today}

\begin{abstract}
In contrast with most Monte Carlo integration algorithms, which are used to estimate ratios, the replica gas identities recently introduced by Adib enable the estimation of absolute integrals and partition functions using multiple copies of a system and normalized transition functions.  Here, an optimized form is presented.  After generalizing a replica gas identity with an arbitrary weighting function, we obtain a functional form that has the minimal asymptotic variance for samples from two replicas and is provably good for a larger number.  This equation is demonstrated to improve the convergence of partition function estimates in a 2D Ising model.
\end{abstract}

\maketitle

%
%

The standard paradigm in Monte Carlo integration is to estimate a ratio.  In the prototypical scenario, a simulation is used to obtain the fraction of randomly sampled points which lie within an encompassing reference region.  The integrand of interest is then estimated by multiplying this fraction with the known reference area.

More sophisticated algorithms retain the feature that they estimate ratios of normalizing constants \cite{Gelman1998} or partition functions \cite{Jarzynski2002}.  Furthermore, the efficiency of all these integration techniques depend on the degree of overlap between the target integrands.  Hence, much effort has been devoted to obtaining intermediate functions which bridge the integrands of interest.

Recently, Adib introduced a class of theorems, which he dubbed the replica gas identities, that can be used to estimate absolute integrals \cite{AdibRepGas}.  These identities are based on the simple physical idea of measuring the volume of a container by filling it with a specified density of ideal gas, and then counting the number of particles.  Similarly, an integrand can be estimated by creating multiple non-interacting copies (or replicas) of a system that fill a region with a specified density, and then counting the number of replicas.  Adib also described a complementary and more abstract identity relevant to simulations with a fixed number of replicas.  A variant of this latter identity, which is particularly applicable to existing parallel tempering (a.k.a. replica exchange) simulations, will be the focus of the present work.

%
%

In this work, we shall consider integrals of the form,
\begin{eqnarray}
Z = \int_\Omega dx \, \pi(x),
\end{eqnarray}
where $\Omega$ is the support of the integral and $\pi(x)$ is a positive-definite function of a $d-$dimensional vector $x$ (see Ref. \cite{AdibRepGas} for a generalization to non-positive definite integrands).  The function $\pi(x)$ can be regarded as an unnormalized probability density.  For many equilibrated physical systems with the energy function $E(x)$, $\pi(x) = e^{-E(x)}$ describes the relative probability of observing $x$, and $Z$ is the partition function.  Discrete integrals can be treated analogously.

Suppose that we have multiple replicas of a system on the same support, each independently sampling from their own respective distributions.  The replica gas identities couple these copies together by means of transition functions, $T(x'|x)$, which are normalized conditional probability distributions that one can sample from and evaluate.  Appropriate transition functions include candidate-generating functions routinely used in Markov Chain Monte Carlo simulations, such as a Gaussian centered about $x$.


For simplicity, we shall first treat the case where the number of replicas is fixed at two.  One replica samples from $\pi(x)$ and the other samples from an auxiliary distribution $\tilde{\pi}(x)$ (which has the auxiliary integral $\tilde{Z} = \int_\Omega dx \, \tilde{\pi}(x)$).  The density $\tilde{\pi}(x)$ is arbitrary and may be $\pi(x)$ itself, but usually corresponds to $\pi(x)$ at a different temperature.  In this case, a generalization of the second replica gas identity \cite{AdibRepGas} is,
\begin{eqnarray}
\label{eq:2RepGas}
Z & = &
\frac{\int_\Omega dx \, [\tilde{\pi}(x)/\tilde{Z}] \int_\Omega dx' \, T(x'|x) \cdot \alpha(x',x) \, \pi(x') }
        {\int_\Omega dx \, [\tilde{\pi}(x)/\tilde{Z}] \int_\Omega dx' \, [\pi(x')/Z] \cdot \alpha(x',x) \, T(x'|x) }
        \nonumber \\ & \equiv & 
\frac{\left< \alpha \, \pi(x') \right>_{\tilde{\pi}, T}}
        {\left< \alpha \, T(x'|x) \right>_{\tilde{\pi},\pi}},
\end{eqnarray}
where the expectation $\left<\mathcal O(x,x')\right>_{f,g}$ of an observable $\mathcal O(x,x')$ means that $x$ and $x'$ are sampled from the distributions $f$ and $g$, respectively.  In the numerator of the equation above, $x$ is sampled from $\tilde{\pi}(x)$, $x'$ is sampled using transition function, and $\pi(x')$ is evaluated.  In the denominator, $x$ is also sampled from $\tilde{\pi}(x)$, but $x'$ is drawn from the replica with density $\pi(x')$, and $T(x'|x)$ is evaluated.  The identity also includes an arbitrary weighting function $\alpha \equiv \alpha(x',x)$; any function is valid as long as the denominator is nonzero.  The choice $\alpha = 1$ reduces the expression to Adib's original identity.

The replica gas identity can be made into an estimator by replacing the expectations with sample means.  While many choices of $\alpha$ will yield statistically consistent expressions, some choices will lead to more efficient estimators than others.  Here, we will optimize $\alpha$ by minimizing the asymptotic variance in estimates of $\zeta \equiv \ln Z$, which we shall denote by $\bar{\zeta}$.  Our procedure will closely follow that of Bennett \cite{Bennett1976}, who minimized the error in an estimator for the \emph{ratio} of partition functions, and yield an analogous estimator.

In the large sample limit, $\bar{\zeta}$ will be Gaussian about the true value of $\zeta$.  Using error propagation based on a first-order Taylor series expansion, the asymptotic variance is,
\begin{eqnarray}
\label{eq:2RepGasVar}
\sigma^2[\bar{\zeta}] & = & 
	\frac{ \left< \alpha^2 \pi(x')^2 \right>_{\tilde{\pi},T} }{ N_T \left< \alpha \, \pi(x') \right>_{\tilde{\pi},T}^2 } + 
	\frac{ \left< \alpha^2 T(x'|x)^2 \right>_{\tilde{\pi},\pi} }{ N_\pi \left< \alpha \, T(x'|x) \right>_{\tilde{\pi},\pi}^2 } 
	- \frac{1}{N_T} - \frac{1}{N_\pi}
\nonumber \\ & = & 
	\frac{ \left< \alpha^2 T(x'|x) \left(\frac{ \pi(x') e^{-\zeta} }{N_T} + \frac{T(x'|x)}{N_\pi} \right) \right>_{\tilde{\pi},\pi} }
	       { \left< \alpha \, T(x'|x) \right>_{\tilde{\pi},\pi}^2 }
	- \frac{1}{N_T} - \frac{1}{N_\pi},
\end{eqnarray}
where $N_T$ and $N_\pi$ are the number of independent samples used to estimate $\left<\mathcal O(x,x')\right>_{\tilde{\pi},T}$ and $\left<\mathcal O(x,x')\right>_{\tilde{\pi},\pi}$, respectively.  The second line is obtained by writing both expectations in terms of $\left<\mathcal O(x,x')\right>_{\tilde{\pi},\pi}$ and combining terms.

As multiplying $\alpha$ by a constant does not change the value of $\sigma^2[\bar{\zeta}]$, we can use a constraint where $\left< \alpha \, T(x'|x) \right>_{\tilde{\pi},\pi}$ (and hence the denominator) is constant.  The variance, Eq. \ref{eq:2RepGasVar}, is then minimized using Lagrange multipliers, leading to the optimal $\alpha$,
\begin{eqnarray}
\label{eq:Opt2Alpha}
\alpha^* \propto \left(\frac{1}{N_T} \pi(x') e^{-\zeta} + \frac{ 1 }{N_\pi} T(x'|x) \right)^{-1}.
\end{eqnarray}
Notably, $\alpha^*$ includes the sought quantity $\zeta$.  The estimator obtained by substituting this function into Eq. \ref{eq:2RepGas} can either be solved by self-consistent iteration, as originally described by Bennett \cite{Bennett1976}, or by finding the zero of an implicit function obtained by rearrangement,
\begin{eqnarray}
\label{eq:Opt2RepGas}
\sum_{l=1}^{N_\pi} \frac{1}{ 1 + \frac{N_\pi}{N_T} \frac{ \pi(x_l') }{ T(x_l'|x_l) } e^{-\bar{\zeta}} } - 
\sum_{m=1}^{N_T} \frac{1}{ 1 + \frac{N_T}{N_\pi} \frac{ T(x_m'|x_m) }{ \pi(x_m') } e^{\bar{\zeta}} } = 0,
\end{eqnarray}
which is similar to an expression described by Shirts et. al. \cite{Shirts2003}.  The variance in $\bar{\zeta}$ in Eq. \ref{eq:Opt2RepGas} may be estimated by substituting the appropriate $\alpha$ into Eq. \ref{eq:2RepGasVar}.

By substituting $\alpha^*$ into Eq. \ref{eq:2RepGasVar}, multiplying by a factor of unity,
\begin{eqnarray}
\frac{1 + \frac{N_T}{N_\pi} \frac{T(x'|x)}{\pi(x')} e^\zeta}{1 + \frac{N_T}{N_\pi} \frac{T(x'|x)}{\pi(x'_l)} e^\zeta},
\end{eqnarray}
and separating the expression into two expectations, we obtain the following formula for the variance,
\begin{eqnarray}
\label{eq:2RepGasOptVar}
\sigma^2[\bar{\zeta}] & = & 
\left[ \sum_{l=1}^{N_\pi} \frac{1}{ 2 + 2 \cosh \left( \zeta \ln \frac{N_T}{N_\pi} \frac{T(x'_l|x_l) }{\pi(x'_l)} \right) } + 
\sum_{m=1}^{N_T} \frac{1}{ 2 + 2 \cosh \left( \zeta \ln \frac{N_T}{N_\pi} \frac{T(x'_m|x_m) }{\pi(x'_m)} \right) } \right]^{-1}
- \frac{1}{N_T} - \frac{1}{N_\pi},
\end{eqnarray}
which is analogous to an extant expression for the error in the Bennett's method \cite{Shirts2003}.


When samples are drawn from $K$ replicas with integrands $\pi_k(x)$ and partition functions $Z_k$ for $k = 1, ..., K$, the following generalized replica gas identity is applicable:
\begin{eqnarray}
\label{eq:RepGas}
Z_t & = & 
\frac{\sum_{k \neq t} \left< \alpha_k \, \pi_t(x') \right>_{\pi_k, T_k}}
       {\sum_{k \neq t} \left< \alpha_k \, T_k(x'|x) \right>_{\pi_k,\pi_t}}.
\end{eqnarray}
Here, $\pi_t(x)$ and $Z_t$ are the targeted integrands and integrals, respectively.  The sums run over all replicas which are not $t$.  As in Eq. \ref{eq:2RepGas}, the choice $\alpha_k = 1$ yields an identity originally described by Adib \cite{AdibRepGas}.

We now consider the optimization of $\alpha$ in the case where $K>2$.  The asymptotic variance of $\bar{\zeta}_t$ is,
\begin{eqnarray}
\label{eq:RepGasVar}
\sigma^2[\bar{\zeta}_t] & = & 
	\frac{\sum_{k \neq t} \frac{1}{N_{T_k}} \left( \left< \alpha_k^2 \pi(x')^2 \right>_{\pi_k,T_k} -  \left< \alpha_k \pi(x') \right>_{\pi_k,T_k}^2 \right)}
	       {\sum_{k \neq t} \left< \alpha_k \pi(x') \right>_{\pi_k,T_k}^2} + 
	\frac{\sum_{k \neq t} \frac{1}{N_{\pi_k}} \left( \left< \alpha_k^2 T_k(x'|x)^2 \right>_{\pi_k,\pi} -  \left< \alpha_k T_k(x'|x) \right>_{\pi_k,\pi_t}^2 \right)}
	       {\sum_{k \neq t} \left< \alpha_k T_k(x'|x) \right>_{\pi_k,\pi_t}^2}
\\ & = &
	\frac{\sum_{k \neq t} \left< \alpha_k^2 T_k(x'|x) \left( \frac{1}{N_{T_k}} \pi(x') e^{-\zeta} + \frac{1}{N_{\pi_k}} T_k(x'|x) \right) \right>_{\pi_k,\pi_t} }
	        { \left[ \sum_{k \neq t} \left< \alpha_k T_k(x'|x) \right>_{\pi_k,\pi_t} \right]^2  } -
	\frac{\sum_{k \neq t} \left(\frac{1}{N_{T_k}} + \frac{1}{N_{\pi_k}} \right) \left< \alpha_k T_k(x'|x) \right>_{\pi_k,\pi_t}^2 }
	        { \left[ \sum_{k \neq t} \left< \alpha_k T_k(x'|x) \right>_{\pi_k,\pi_t} \right]^2 }
\end{eqnarray}
Unlike in the case where $K=2$, the second term in the above variance expression does not reduce to a constant.  We can show, however, using a procedure similar to Guibas and Veach \cite{Veach1995, Veach1997}, that it is bounded by a constant:
\begin{eqnarray}
& & \frac{\sum_{k \neq t} \left(\frac{1}{N_{T_k}} + \frac{1}{N_{\pi_k}} \right) \left< \alpha_k T_k(x'|x) \right>_{\pi_k,\pi_t}^2 }
	        { \left[ \sum_{k \neq t} \left< \alpha_k T_k(x'|x) \right>_{\pi_k,\pi_t} \right]^2 }
\nonumber \\ & \leq & 
\frac{ max \left(\frac{1}{N_{T_k}} + \frac{1}{N_{\pi_k}} \right) \sum_{k \neq t} \left< \alpha_k T_k(x'|x) \right>_{\pi_k,\pi_t}^2 }
	        { \left[ \sum_{k \neq t} \left< \alpha_k T_k(x'|x) \right>_{\pi_k,\pi_t} \right]^2 }
\nonumber \\ & \leq & 
\frac{ max \left(\frac{1}{N_{T_k}} + \frac{1}{N_{\pi_k}} \right) \left[ \sum_{k \neq t} \left< \alpha_k T_k(x'|x) \right>_{\pi_k,\pi_t} \right]^2 }
	        { \left[ \sum_{k \neq t} \left< \alpha_k T_k(x'|x) \right>_{\pi_k,\pi_t} \right]^2 }
\nonumber \\ & = & 
max \left(\frac{1}{N_{T_k}} + \frac{1}{N_{\pi_k}} \right)
\end{eqnarray}
Thus, if we find a set of functions for $\alpha_k$ that minimize the first term in $\sigma[\bar{\zeta}_t]$, no other choices for $\alpha_k$ can decrease the variance by more than $max \left(\frac{1}{N_{T_k}} + \frac{1}{N_{\pi_k}} \right)$.

Minimization of the first term in Eq. \ref{eq:RepGasVar} by Lagrange multipliers leads to an optimized set of $\alpha_k$ analogous to Eq. \ref{eq:Opt2Alpha}.  As with $K=2$, substitution of this $\alpha_k$ leads to an estimator for the integral which can either be solved self-consistently or by finding the zero of an implicit function,
\begin{eqnarray}
\label{eq:OptRepGas}
\sum_{k \neq t} \left[ 
\sum_{l=1}^{N_{\pi_k}} \frac{1}{ 1 + \frac{N_{\pi_k}}{N_{T_k}} \frac{ \pi(x_l') }{ T_k(x_l'|x_l) } e^{-\bar{\zeta}} } - 
\sum_{m=1}^{N_{T_k}} \frac{1}{ 1 + \frac{N_{T_k}}{N_{\pi_k}} \frac{ T_k(x_m'|x_m) }{ \pi(x_m') } e^{\bar{\zeta}} } \right] = 0,
\end{eqnarray}
The corresponding variance may be estimated by substituting the appropriate $\alpha_k$ into Eq. \ref{eq:RepGasVar}.

%
%

The original and optimized forms of the replica gas identity were demonstrated by estimating the partition function of a two-dimensional $32 \times 32$ spin Ising model, using parameters similar to those previously described \cite{AdibRepGas}.  As in a typical Ising model simulation without magnetization, the energy of a spin configuration $x$ was given by $E(x) = -\sum_{<k,l>} x_k x_l$, a sum over all pairs of neighbors \cite{Krauth2006}.  Periodic boundary conditions were used.  Simulations were performed for 26 replicas at the temperatures $\beta^{-1} = 0.5, 0.6, ..., 3.0$.  Configurations were equilibrated using 250 iterations in which a cluster move, Woolf's generalization \cite{Wolff1989} of the Swendsen-Wang algorithm \cite{SwendsenWang1987}, was followed by attempted replica exchanges between adjacent temperature pairs (alternating between $\{ (\beta_1, \beta_2), (\beta_3, \beta_4), ..., (\beta_{K-1},\beta_K) \}$ and $\{ (\beta_2, \beta_3), (\beta_4, \beta_5), ..., (\beta_{K-2},\beta_{K-1}) \}$).  Equilibration was followed by $10^5$ production steps using Monte Carlo trial moves generated by a transition function $T(x'|x)$ that flips each spin with probability $p_{flip} = 1/N$, where $N$ is the total number of spins.  The transition function was the same for every replica.  After every 100 steps of production, a replica exchange was attempted between a random replica and its higher-temperature neighbor, energies of each configuration were stored, $T(x'|x)$ was used to sample configurations from $\left< \mathcal O(x,x') \right>_{\pi_k, T}$ (these are distinct from the Markov chain), and transition probabilities between the different configurations were calculated using $T(x'|x) = (p_{flip})^{||x'-x||} (1-p_{flip})^{N - ||x'-x||}$, where $||x'-x||$ is the number of spin mismatches \cite{AdibRepGas}.

\begin{figure}[h]
\begin{center}
\includegraphics{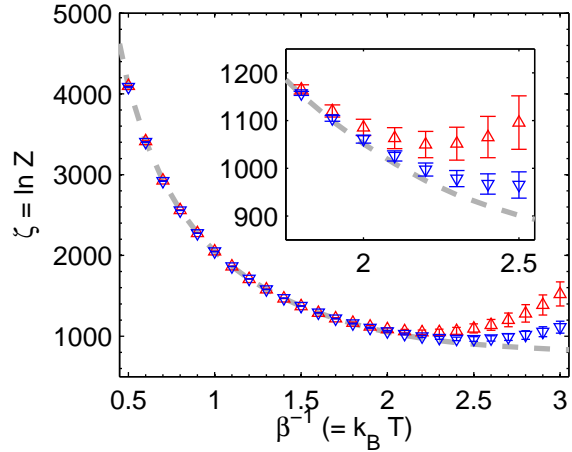}
\caption{\label{fig:Ising2D}
Comparison of logarithmic partition function estimates: the exact Kaufman formula (dashed line), the original replica gas identity, Eq. \ref{eq:RepGas} with $\alpha = 1$ (upward triangles), and the optimized form, Eq. \ref{eq:OptRepGas} (downward triangles).  Markers and error bars indicate the mean and standard deviation of 1000 independent simulations.  Inset: magnification around the critical temperature.
}
\end{center}
\end{figure}

For this system, we find that both estimators have similar and relatively accurate efficiencies at lower temperatures (Fig. \ref{fig:Ising2D}).  Closer to and above the critical temperature ($k_B T = 2.269$), however, the optimized estimator (Eq. \ref{eq:OptRepGas}) has less variance and bias than the original estimator (Eq. \ref{eq:RepGas}) with unit weight.  Nonetheless, the high-temperature estimates are substantially worse than those at low temperatures.

Ultimately, the quality of the optimized expression is subject to the same fundamental limitations as Adib described with the original estimator \cite{AdibRepGas}.  For the convergence of averages to not be dominated by rare events, (a) most configurations generated by the transition function should fall in typical regions of phase space for the target distribution and (b) most pairs of configurations from different replicas should have substantial density in $T(x'|x)$.  These requirements hold true in the low-temperature regime, where the spins are relatively ordered and a local transition function suffices, but are not met at higher temperatures, where configurations have disordered spins.  As Adib mentioned, convergence may be improved by fine-tuning the simulation procedure or choosing a more suitable transition function.

Overcoming this sampling limitation is not the intent of the current work.  Rather, the purpose is to improve the analysis of a given data set.  While the resulting expression (Eq. \ref{eq:OptRepGas}) is more complex, the increased difficulty of implementation is no greater than using the Bennett Acceptance Ratio \cite{Bennett1976} relative to unidirectional estimates of free energy \emph{differences}, and does not require sampling from more ensembles.  Expected statistical efficiency gains should make this investment worthwhile.

\section{Acknowledgments}

D.M. thanks Artur Adib for sharing and explaining his manuscript, and Lee Makowski for sponsoring a Director's Postdoctoral Fellowship.


\end{document}